\documentclass[draft,finalnew]{agujournal2019}
\usepackage{color}
\usepackage{graphicx}
\usepackage{amsmath}
\usepackage{amssymb}
\usepackage{cancel}
\usepackage{apacite}
\usepackage{natbib}
\usepackage[symbol]{footmisc}
\draftfalse
\graphicspath{{./figs/}}
\makeatletter
\def\input@path{{./figs/}}
\makeatother

\def\sech{\text{sech}}

\renewcommand{\vec}[1]{\mathbf{#1}}

\newcommand{\ddt}[2]{\frac{\partial{#1}}{\partial{#2}}}

\newcommand{\comment}[1]{}
\newcommand{\ig}[2]{\includegraphics[width = #1]{#2}}

\journalname{JGR: Space Physics}
\begin{document}
\title{Drift instabilities in thin current sheets using a two-fluid model with pressure tensor effects}
\authors{Jonathan Ng \affil{1}\footnote{Current affiliation: Department of Astronomy, University of Maryland, College Park, MD 20742, USA}, Ammar Hakim \affil{2}, J. Juno\affil{3} and A. Bhattacharjee \affil{1,2}}

\affiliation{1}{Department of Astrophysical Sciences, Princeton University, Princeton, NJ 08543, USA}
\affiliation{2}{Princeton Plasma Physics Laboratory, Princeton, NJ 08544, USA}
\affiliation{3}{Institute for Research in Electronics and Applied Physics, University of Maryland, College Park, MD 20742, USA}
\date{\today}

\begin{keypoints}
\item The drift kink and lower hybrid drift instabilities are studied using a \change{ten moment model}{ten-moment fluid model}.
\item Inclusion of the non-gyrotropic pressure tensor improves agreement with kinetic results for the kink mode
\item Ion physics of the lower hybrid drift instability can be reproduced using a nonlocal heat flux closure.
\end{keypoints}

\begin{abstract}
The integration of kinetic effects in fluid models is important for global simulations of the Earth's magnetosphere. We use a two-fluid ten moment model, which includes the pressure tensor and has been used to study reconnection, to study the drift kink and lower hybrid drift instabilities. Using a nonlocal linear eigenmode analysis, we find that for the kink mode, the ten moment model shows good agreement with kinetic calculations with the same closure model used in reconnection simulations, while the electromagnetic and electrostatic lower hybrid instabilities require modeling the effects of the ion resonance using a Landau fluid closure. Comparisons with kinetic simulations and the implications of the results for global magnetospheric simulations are discussed.
\end{abstract}

\section{Introduction}

Thin current sheets are often found in the Earth's magnetosphere, and are unstable to a variety of modes, including the tearing mode, drift-kink mode and lower hybrid drift instability (LHDI). 

The drift-kink mode is an ion scale mode ($k\rho_i \sim 1$) driven by the streaming of ions and electron, and was once thought to be a possible mechanism for substorm onset, becoming the subject of theoretical and numerical studies using fluid and kinetic theory \citep{daughton:1999kink,daughton:1998,pritchett:1996,zhu:1996,yoon:1998,ozaki:1996}\change{citation to Ozaki added}{}. However, it was shown \citep{daughton:1999} that the electron-ion drift-kink instability has a strongly reduced growth rate at the physical mass ratio. More recently, there has been work on ion-ion kink instabilities driven by the velocity difference between background and current carrying ions \citep{karimabadi:2003a,karimabadi:2003b}. 

Compared to the drift kink instability, the electrostatic lower hybrid drift instability (LHDI) has shorter wavelength, with a broad range of wavenumbers  $(m_e/m_i)^{1/4} < k\rho_e < 1$ with frequency $\omega \approx \Omega_{lh} \sim \sqrt{\Omega_{ce}\Omega_{ci}}$ \citep{daughton:1999,davidson:1977}. These fluctuations are located at the edge of the current sheet, where the density gradient is strongest, and have been observed in space, experiments and simulations \citep{bale:2002, carter:2002, lapenta:2003}. While the electrostatic LHDI does not always enhance \remove{always} reconnection by itself due to its location away from the centre of the current sheet, it can alter the structure of the current layer due to its comparatively faster growth rate and drive secondary instabilities such as the drift-kink or Kelvin-Helmholtz instabilities \citep{price:2016, lapenta:2003, daughton:2003}. Additionally, there is also a longer wavelength $k\sqrt{\rho_e\rho_i} \sim 1$ electromagnetic lower hybrid mode which has a lower growth rate \citep{daughton:2003}. This instability can be observed at the centre of the current sheet and can influence the reconnection process \citep{roytershteyn:2012}. Within the magnetosphere, there have been observations of the LHDI at both the magnetopause \citep{graham:2017} and magnetotail \citep{zhou:2009}. 

The Earth's magnetosphere is comprised mostly of a collisionless plasma. Global simulations of the magnetosphere have relied so far mostly on single-fluid MHD, which is inadequate for collisionless plasmas. In recent years, we have attempted to extend fluid models to incorporate more kinetic effects in magnetospheric systems using higher moment models \citep{wang:2018}. While these models have been successful in simulating large reconnecting systems \citep{wang:2015, wang:2018, ng:2015, ng:2017, allmann:2018}, there have not been detailed studies on how well the drift instabilities are represented by the models. Though there is some work on these instabilities in field-reversed configurations \citep{hakim:2007}, existing fluid theory for current sheets \citep{daughton:1999kink,yoon:2002,yoon:1998, pritchett:1996} shows some discrepancies with kinetic theory \citep{davidson:1977,daughton:1999,daughton:2003}, and does not include the pressure tensor which is evolved by the ten moment model. 

In the light of these attempts, it is important to understand if the extended fluid equations can model these instabilities, and if inclusion of the pressure tensor and associated closure improves the agreement between kinetic and fluid models. It is also necessary to determine if the same closures which give good agreement with kinetic studies of reconnection can simultaneously describe the instabilities. One area of interest is the growth rate of the kink and sausage modes, where the fluid calculations can have faster growth rates at shorter wavelengths \citep{daughton:1999kink, yoon:2002, pritchett:1996}, while in kinetic theory, the fastest growing kink mode is around $k L  \sim 1$, $L \sim \rho_i$ being the length scale of gradients in the equilibrium, and the sausage mode is stable \citep{daughton:1999}. 

This paper is focused on linear eigenmode calculations of the drift instabilities in Harris sheets \citep{harris:1962} using the five and ten moment models. The five moment model is a standard two fluid model with isotropic pressure, and reduces to Hall MHD in the limit of $m_e\to 0$, $c\to \infty$ and $n_i = n_e$, while the ten moment model includes the effects of an anisotropic pressure tensor and a heat flux closure. Our results show that the ten moment model is able to model the drift kink instability and magnetic reconnection simultaneously, while a proper treatment of the lower hybrid instabilities requires capturing the ion kinetic response using a Landau fluid closure for the heat flux, though the instability still appears when using a simple fluid model. The remainder of the paper is organised as follows: Section \ref{sec:eqns} describes the moment models and closures used in the calculations, and Section~\ref{sec:eigenmode} describes the linear eigenmode calculations. The results of the kink and LHDI calculations are shown in Sections~\ref{sec:kink} and~\ref{sec:lhdi}, with some discussion of the appropriate closure to use for the LHDI in Section~\ref{sec:lhdi}. Finally, comparisons between fluid and a fully kinetic Vlasov-Maxwell simulation are presented in Section~\ref{sec:simulations}, and we conclude in Section~\ref{sec:conclusion}.

\section{Moment equations}
\label{sec:eqns}

For each species, the fluid equations are obtained by taking velocity moments of the Vlasov equations. This leads to 
\begin{equation}
\begin{split}
\ddt{n}{t} + \ddt{}{x_j}(nu_j) &= 0\\
m \ddt{}{t}(nu_i) + \ddt{\mathcal{P}_{ij}}{x_j} &= nq(E_j + \epsilon_{ijk}u_jB_k). \\
\end{split}
\end{equation}
where $\mathcal{P}_{ij}$ is the second moment of the distribution function
\begin{equation}
\mathcal{P}_{ij} \equiv m\int v_iv_j f d^3v
\end{equation}
In the five moment model, the pressure is assumed to be isotropic, and we evolve the energy equation in addition to the continuity and momentum equations. 
\begin{equation}
\ddt{\mathcal{E}}{t} + \ddt{}{x_j}\left(u_j\left(p + \mathcal{E}\right)\right) = n q u_i E_i. 
\end{equation}
Here $\mathcal{E} = \tfrac{1}{2}n m u^2 + nm\epsilon$, where $\epsilon = P/[(\gamma-1)n m]$ is the internal energy per unit mass. For this paper we use $\gamma = 5/3$. 

The ten moment model evolves the full pressure tensor according to
\begin{equation}
\ddt{\mathcal{P}_{ij}}{t} + \ddt{\mathcal{Q}_{ijk}}{x_k} = nqu_{[i}E_{j]} + \frac{q}{m}\epsilon_{[ikl}\mathcal{P}_{kj]}B_l,
\end{equation}
where $\mathcal{Q}_{ijk}$ is the third moment of the distribution function 
\begin{equation}
\mathcal{Q}_{ijk} \equiv m\int v_iv_jv_k f d^3v,
\end{equation}
and the square brackets denote a sum over permutations of the indices (e.g.~$u_{[i}E_{j]} = u_iE_j + u_jE_i$). Following \citep{wang:2015} one can write $\mathcal{Q}_{ijk}$ in terms of the heat flux tensor $Q_{ijk} \equiv m\int (v_i-u_i)(v_j-u_j)(v_k-u_k) f d^3v$
\begin{equation}
\mathcal{Q}_{ijk} = Q_{ijk} + u_{[i}\mathcal{P}_{jk]} - 2 m n u_iu_ju_k.
\end{equation}

For collisionless plasmas in the unmagnetised limit, we use a three-dimensional extension of the Hammett-Perkins closure, which can be expressed as follows for both electrons and ions \citep{hammett:1990}:
\begin{equation}
q_{ijk}(\vec{x}) = n(\vec{x})\hat{q}_{ijk}(\vec{x})
\end{equation}
where $\hat{q}_{ijk}$ in Fourier space is $\tilde{q}_{ijk}$ and is calculated as
\begin{equation}
\tilde{q}_{ijk} = -i \frac{v_{t}}{|k|}\chi k_{[i}\tilde{T}_{jk]}.
\label{eq:closure}
\end{equation}
Here $\tilde{T}_{jk}$ is the Fourier transform of the deviation of the local temperature tensor from the mean. The $1/|k|$ scaling makes this a non-local closure when expressed in real space \citep{hammett:1992,snyder:2001} and provides a 1 to 3 pole Pad\'e approximation of various components of the dielectric tensor. The coefficient $\chi = \sqrt{4/9\pi}$ is the best fit value for the diagonal $q_{iii}$ component and reduces to the closure in \citep{hammett:1990,hammett:1992} in the 1-D limit. This closure has been used to study reconnection in the context of magnetic island coalescence, and gives better agreement with kinetic results than Hall MHD \citep{ng:2017}. 

Due to the computational costs involved in calculating the nonlocal heat flux, relaxation of the pressure tensor to local isotropy is a more common approximation, and has been used successfully in large scale studies of reconnection and magnetospheres \citep{wang:2015,ng:2015,wang:2018}. With this model the heat flux divergence term is replaced by $\partial_iQ_{ijk} =  v_{t}|k_{0}|(P_{ij}-P\delta_{ij})$ \citep{wang:2015,ng:2015,hesse:1995,yin:2001}, where $v_t = \sqrt{2 T/m}$ is the thermal velocity of the associated species and $k_{0}$ is a free parameter for each species. 

As this work is focused on understanding if the drift instabilities exist within the ten-moment model and whether they will be present in global simulations and interact with reconnection, we study both local relaxation and the nonlocal closure over a variety of parameter regimes. 

\section{Eigenmode calculations}
\label{sec:eigenmode}

To study the current sheet instabilities, we begin with the exact Harris equilibrium \citep{harris:1962}. The magnetic field and density are described by
\begin{align}
B_x(z) &= B_0 \tanh\left(\frac{z}{L}\right)\\
n(z) &= n_0 \, \sech^2\left(\frac{z}{L}\right),
\end{align}
with species drift velocities 
\begin{equation}
u_{y,s} = \frac{2 \remove{c} T_s}{q_s B_0 L}
\end{equation}
The temperature is determined by the equilibrium condition \change{$\beta = 1$}{$\beta_e+\beta_i=1$. Here $\beta_s$ is the species plasma beta defined as $2\mu_0 n_0 T_s/B_0^2$. } 

We consider perturbations about the equilibrium in the form $f(y,z,t) = f_1(z)\exp\left(i \left(k_y y - \omega t\right)\right)$, with no variation in the $x$ direction (parallel to the equilibrium magnetic field). This is orthogonal to the usual $2$-D plane used in reconnection studies. For the modes we are studying, the perturbed quantities $B_y, B_z, E_x$ and $v_{x}$  are identically zero \citep{pritchett:1996}, and in the ten-moment model, the pressure tensor components $P_{xy}$ and $P_{xz}$ are also zero. This leads to reduced systems of $11$ and $17$ equations for the five and ten-moment models respectively (for two species).

For the five moment equations, they are (normalised to $c=1$, $d_{i0} = 1$, $\omega_{pi0} = 1$ \add{in simulation units}):
\begin{equation}
\begin{split}
(\omega - k_y u_s) n_{1,s} - k_y n_0 v_{1y,s} + i (n_0' v_{1z,s} + n_{0,s} v_{1z,s}') = 0\\
(\omega - k_y u_s) v_{1y,s} - \frac{i q_s}{m_s} E_y - i\Omega_s v_{1z,s} - \frac{k_yP_{1,s}}{n_0m_s}= 0\\
(\omega - k_y u_s) v_{1z,s} - \frac{i q_s}{m_s} E_z + i\Omega_s v_{1y,s} + i \frac{q_s v_s}{m_s} B_{1x} \\ 
+ i \Omega_s\frac{ n_{1,s}}{n_0} u_s + i \frac{P_{1,s}'}{m_sn_0} = 0\\
(\omega - k_y u_s) P_{1,s} + i v_{1z,s} P_{0,s}' - \gamma k_y P_0 v_{1y,s} + i \gamma P_{0,s} v_{1z,s}'= 0\\
\omega B_{1x} - (k_y E_z + i E_y') = 0\\
\omega E_y - i B_{1x}' + i\sum_s  q_s (n_0 v_{1y,s} + n_{1s} v_s) = 0\\
\omega E_z -  k_yB_{1x} + i \sum_s q_s n_0 v_{1z,s} = 0,\\
\label{eq:fivemom}
\end{split}
\end{equation}
where the primes represent $z$ derivatives and $\Omega_s = q_sB_0(z)/m_s$.

The linear ten-moment equations are as follows: 
\begin{equation}
\begin{split}
(\omega - k_y u_s) n_{1,s} - k_y n_0 v_{1y,s} + i (n_0' v_{1z,s} + n_{0,s} v_{1z,s}') = 0\\
(\omega - k_y u_s) v_{1y,s} - \frac{i q_s}{m_s} E_y - i\Omega_s v_{1z,s} - \frac{k_yP_{1yy,s}}{n_0m_s} + i \frac{P_{1yz,s}'}{n_0m_s}= 0\\
(\omega - k_y u_s) v_{1z,s} - \frac{i q_s}{m_s} E_z + i\Omega_s v_{1y,s} + i \frac{q_s u_s}{m_s} B_{1x} + i \Omega_s\frac{ n_{1,s}}{n_0} u_s - k_y\frac{P_{1yz,s}}{n_0m_s} + i \frac{P_{1zz,s}'}{m_sn_0} = 0\\
(\omega - k_y u_s) P_{1xx,s} - k_y P_{0,s} v_{1y,s} + i P_{0,s} v_{1z,s}' + i P_{0,s}' v_{1z,s} + i |k_{0,s}| v_{t,s} \left(P_{1xx,s} - \frac{P_{1,s}}{3}\right) = 0\\
(\omega - k_y u_s) P_{1yy,s} + i P_{0,s} (i k_y v_{1y,s} + v_{1z,s}') - 2 k_y P_{0,s} v_{1y,s} + i v_{1z,s} P_{0,s}' - 2 i \Omega_j P_{1yz,j} + i |k_{0,s}| v_{t,s}\left(P_{1yy,s} -\frac{P_{1,s}}{3}\right) = 0\\ 
(\omega -k_y u_s) P_{1yz,s} + i P_{0,s} v_{1y,s}' - k_y P_{0,s} v_{1z,s} - i \Omega_s (P_{1zz,s}- P_{1yy,s}) + i |k_{0,s}| v_{t,s} P_{1yz,s} = 0\\
(\omega -k_y u_s) P_{1zz,s} + i P_{0,s} (i k_y v_{1y,s} + v_{1z,s}') + 2 i P_{0,s} v_{1z,s}' + i v_{1z,s} P_{0,s}' + 2 i \Omega_s P_{1yz,s} + i |k_{0,s}| v_{t,s} \left(P_{1zz,s} - \frac{P_{1,s}}{3}\right) = 0,\\
\label{eq:tenmom}
\end{split}
\end{equation}
where $P_1 = P_{1xx} + P_{1yy} + P_{1zz}$ is the perturbed trace of the pressure tensor and Maxwell's equations remainin the same. The modifications are the additional equations for the pressure tensor components and the replacement of the pressure gradient by the divergence of the pressure tensor in the momentum equations. The terms proportional to $|k_{0,s}|$ in the pressure tensor evolution represent the local isotropisation discussed in Section~\ref{sec:eqns}. 

When using the nonlocal closure, we replace the relaxation terms in Eq.~\eqref{eq:tenmom} with the following expressions for the nonlocal heat flux
\begin{equation}
\begin{split}
i \left(\nabla\cdot\vec{q}\right)_{xx} &= i \sqrt{\frac{4}{9\pi}}k_y v_t (P_{1,xx} - n_1 T_0) \\
i \left(\nabla \cdot \vec{q} \right)_{yy} &= i \left(\sqrt{\frac{4}{\pi}}k_y v_t (P_{1,yy} - n_1 T_0) \right) + \frac{2}{3}\sqrt{\frac{4}{\pi}}v_t \ddt{P_{1,yz}}{z} \\
i \left(\nabla\cdot\vec{q}\right)_{yz} &= i\left( \frac{2}{3}\sqrt{\frac{4}{\pi}}k_y v_t P_{1,yz}\right) + \frac{1}{3}\sqrt{\frac{4}{\pi}}v_t \ddt{(P_{1,zz}-n_1 T_0)}{z}\\
i \left(\nabla\cdot\vec{q}\right)_{zz} &= i \left( \sqrt{\frac{4}{9\pi}}k_y v_t (P_{1,zz} - n_1 T_0)\right)
\end{split}
\end{equation}
Here we have only kept the $k_y$ terms in $q_{ijk} \propto k_{[i}T_{jk]}/|k|$. 

Although it is possible to reduce the five-moment system to a single second-order differential equation which is amenable for analysis \citep{yoon:2002}, the additional equations in the ten-moment system make it somewhat difficult to use the same method. Instead, we note that the equations can be written as 
\begin{equation}
\omega \vec{F} + k_y \mathcal{A}_y \vec{F} + \mathcal{A}_z \ddt{\vec{F}}{z} +\mathcal{S} \vec{F} = \vec{0}, 
\label{eq:eigenmode}
\end{equation}
where $\mathcal{A}_y, \mathcal{A}_z$ and $\mathcal{S}$ are coefficient matrices. The instabilities of the system can then be found directly by discretizing the equations and solving for the eigenvalues of the resulting matrix. In this work we used 6th order central differences to calculate the $z$ derivatives. The equations are solved from $z = -12.8L$ to $z = 12.8 L$. The resolution of kink modes and longer wavelength lower hybrid modes typically requires fewer than 250 grid points. For shorter wavelength lower hybrid modes, which are more localised and can have finer structure, we use a smaller domain $z = -6.8 L$ to $z = 6.8 L$, and 250 points. Once specific eigenvalues are found, convergence is tested by increasing resolution by a factor of four and using a sparse solver to find the closest solutions to the selected eigenvalue. 

One feature of this method compared to the search methods employed by \citep{daughton:1999,yoon:2002} is that we find \textit{all} the modes of the system (limited by the resolution and numerical method), and post-processing is necessary to identify the unstable modes of interest. 

\section{Drift-kink instability}
\label{sec:kink}
The solution of Eq.~\eqref{eq:eigenmode} for both systems leads to a spectrum of eigenmodes over a range of $k_y$. In this -- and the following -- section we compare the five and ten moment solutions for the drift-kink and lower hybrid instabilities. Where possible, we use similar parameters to the kinetic calculations in the literature \citep{daughton:1999,daughton:2003}.

\begin{figure}
\ig{3.375in}{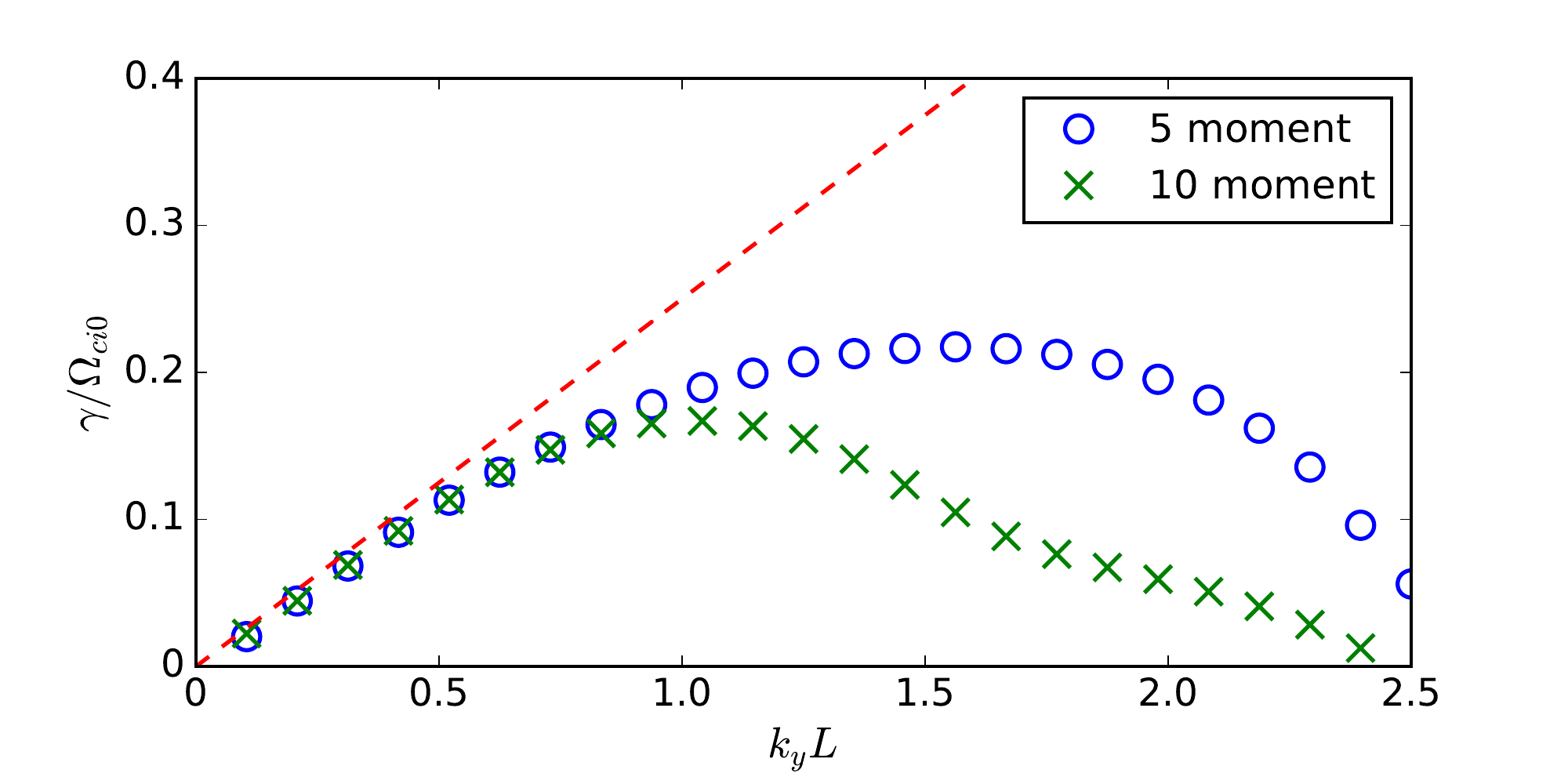}
\caption{Growth rate of the drift-kink instability in an electron-positron plasma.}
\label{fig:ep-plasma}
\end{figure}

We begin by studying the case of an electron-positron plasma, $m_i/m_e = 1, T_i/T_e = 1, \rho_i/L = 0.5, v_{t,e} = 0.25 c$. This particular set of parameters has been studied in earlier work \citep{daughton:1999kink,pritchett:1996} and is a useful basis for direct comparison. Figure~\ref{fig:ep-plasma} shows the fastest growing kink modes for this configuration. \change{The five moment results are comparable to those of}{The variation of the growth rate with $k_y$ shown in Fig.}~\ref{fig:ep-plasma} \add{shows good agreement with the results of} \citep{pritchett:1996}, with a maximum growth rate of $\gamma/\Omega_{ci} = 0.22$, while the ten-moment result shows a maximum of $\gamma/\Omega_{ci} = 0.17$ at a longer wavelength with $k_y L \approx 1$. This is in better agreement with the linear Vlasov results in \citep{daughton:1998,daughton:1999kink}. With the ten-moment model, there is a plateau for $k_y L > 1.5$, which is sensitive to the value of $k_{0,i}$ used. For this set of results we used \add{local relaxation with} $k_{0,e} = 1/d_e$, $k_{0,i} = 1/d_i$, a choice similar to that used in earlier reconnection studies \citep{ng:2015, wang:2015}. 

At long wavelengths, both models approach the dashed lines, which show the incompressible solution \citep{daughton:1999kink}
\begin{align}
\omega_r &= \frac{k_y u_i}{1+m_e/m_i}\left(1 - \frac{T_e m_e}{T_i m_i}\right) \\
\gamma &= \frac{k_y u_i}{1+m_e/m_i}\left(\frac{m_e}{m_i}\right)^{1/2} \left(1+ \frac{T_e}{T_i}\right).
\end{align}

\begin{figure}
\ig{3.375in}{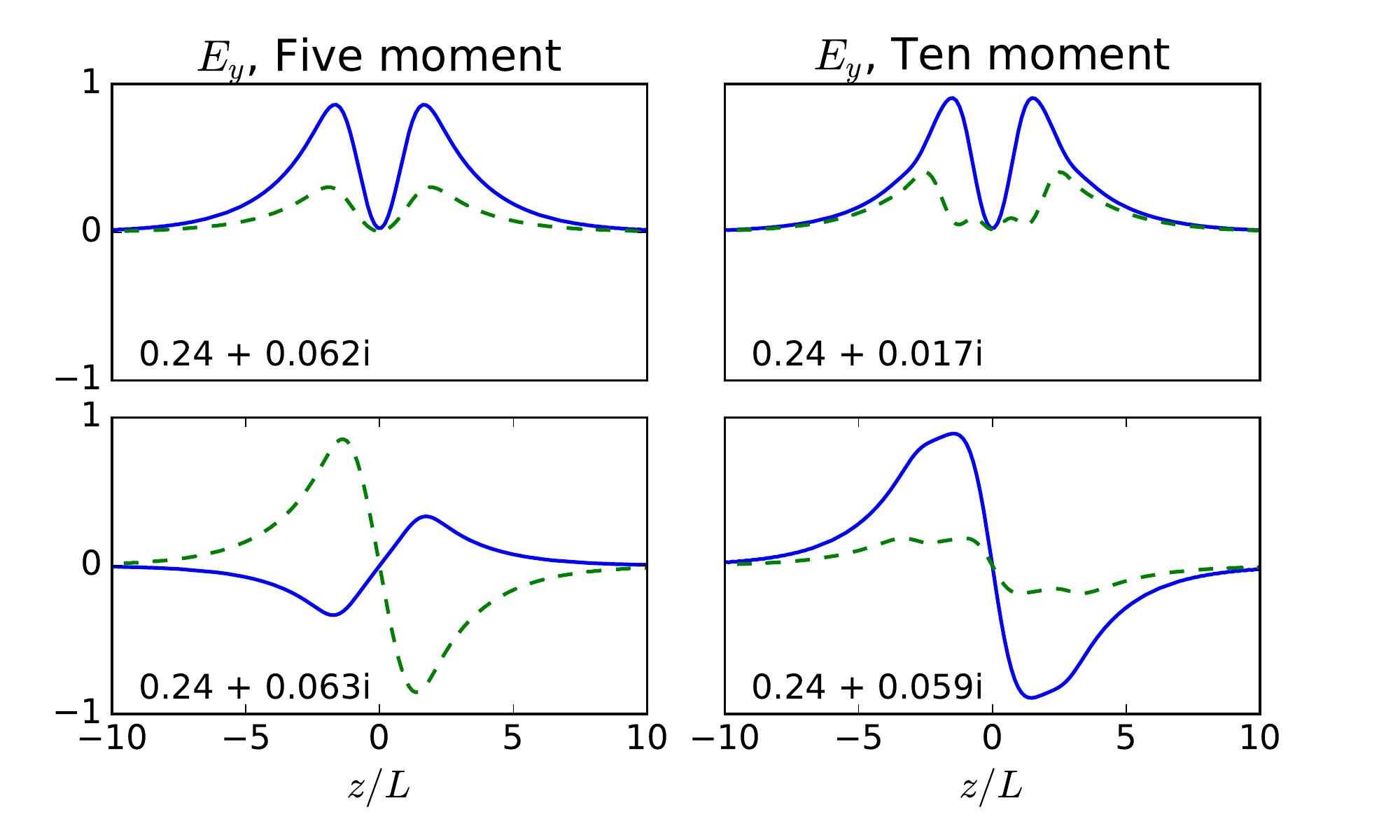}
\caption{Real (solid) and imaginary (dashed) parts of $E_y$ for five and ten moment models. The real frequency and growth rate are normalised to $\Omega_{ci0}$. . \add{The top row shows sausage modes with even $E_y$ profiles, while the bottom row shows kink modes with odd $E_y$ profiles.} }
\label{fig:kink_modes}
\end{figure}

The equation systems \eqref{eq:fivemom} and \eqref{eq:tenmom} support a spectrum of eigenmodes. In Fig.~\ref{fig:kink_modes} we show the mode structure of unstable odd and even harmonics for both five and ten moment models at a fixed wavenumber $k_yL = 0.5$. Other physical parameters are $m_i/m_e = 64, \rho_i/L = 0.7$ and $T_i/T_e = 1$. In the left column, the five moment eigenfunctions are shown, with both odd and even (kink and sausage) modes supported by the system. The real frequencies are consistent with the ion diamagnetic frequency, with $k_yu_i/\Omega_{ci0} = 0.245$. In the right column, the ten moment eigenmodes are shown. We were only able to find a single kink mode growing at a similar growth rate to the five moment solutions, with the sausage mode growth rate more than a factor of three smaller. 

\begin{figure}
\ig{3.375in}{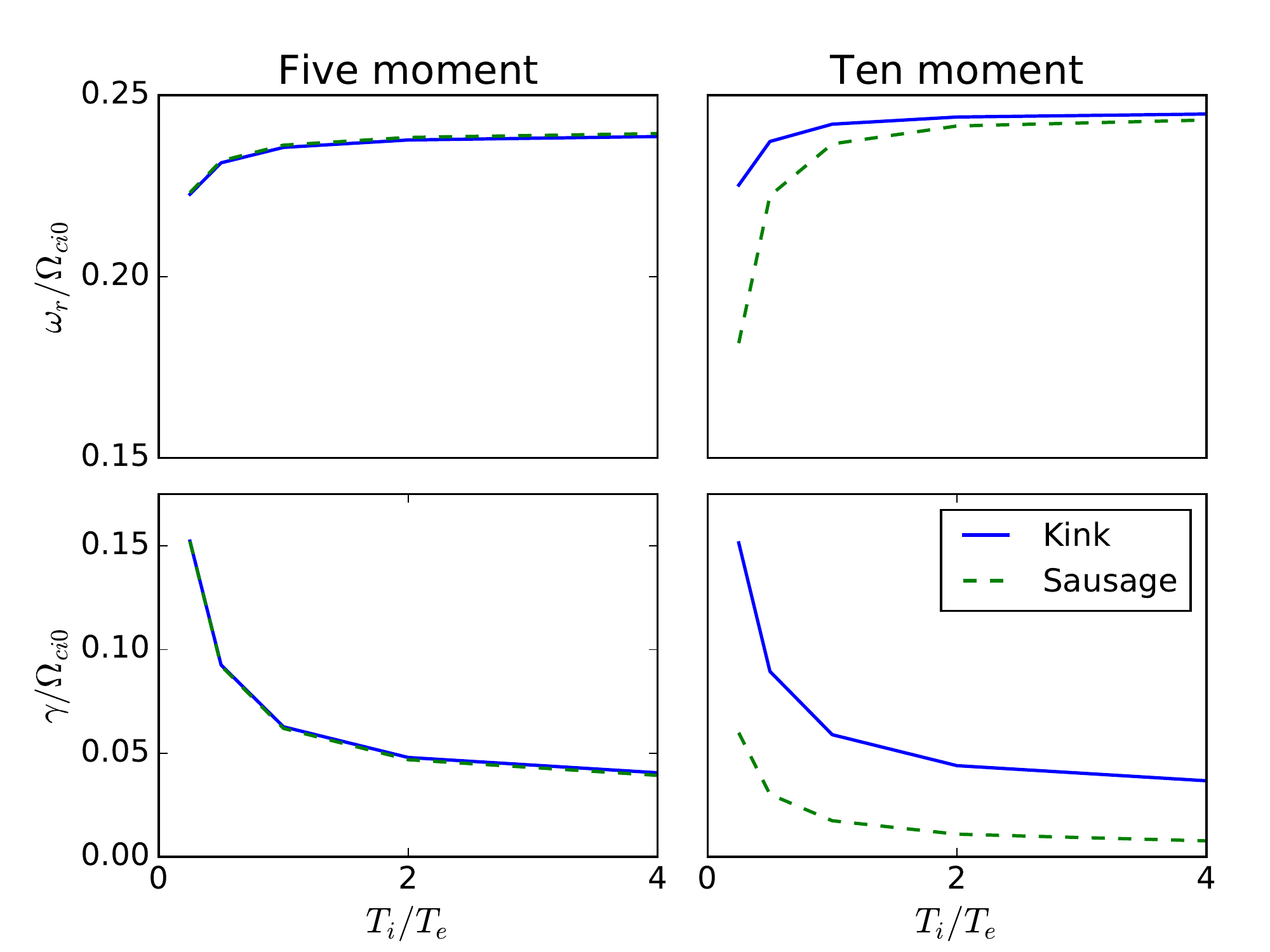}
\caption{Normalised growth rate and real frequencies of the kink and sausage modes using the five and ten moment models as a function of the temperature ratios. Parameters are $m_i/m_e = 64, \rho_i/L = 0.7$ and $k_yL = 0.5$. }
\label{fig:tite}
\end{figure}

\begin{figure}
\ig{3.375in}{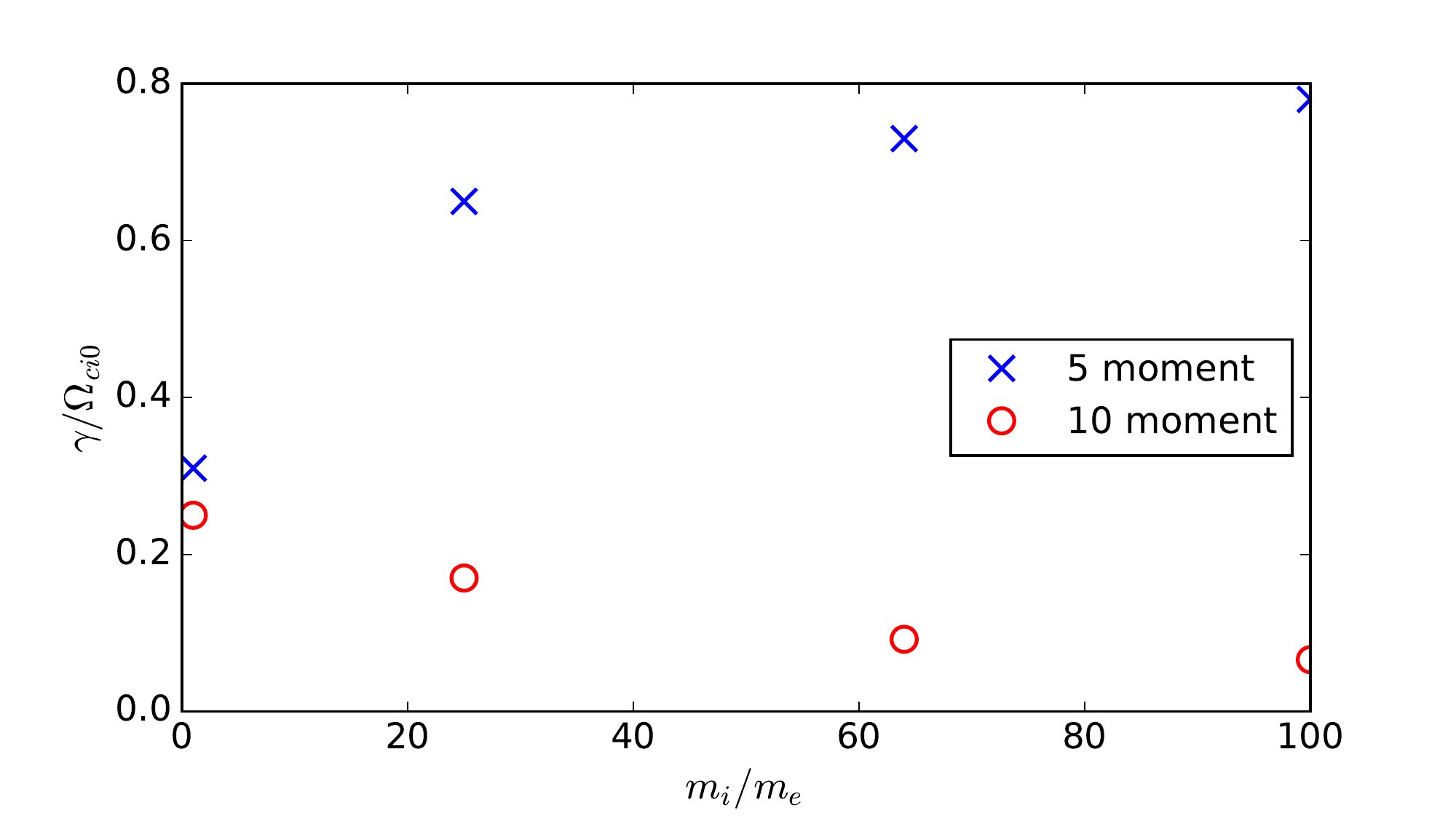}
\caption{Fastest growth rate as a function of the ion/electron mass ratio. }
\label{fig:mime}
\end{figure}

The scaling of the models with physical parameters is shown in Figs.~\ref{fig:tite} and \ref{fig:mime}. In Fig.~\ref{fig:tite} the scaling of the growth rates and frequencies of the kink and sausage modes with the ratio of ion and electron temperatures is shown. Here we are comparing modes with structure similar to those shown in Fig.~\ref{fig:kink_modes}. In both models, the growth rates increase as $T_i/T_e$ decreases, in agreement with kinetic calculations \citep{daughton:1999}. The differences between the models are evident in the sausage mode growth rates, where the five moment model shows a sausage mode growing at almost the same rate as the kink mode, similar to \citep{pritchett:1996}, while the sausage mode in the ten-moment model grows 3 to 4 times more slowly than the kink mode across the range of temperatures. 

In large scale simulations, the use of a reduced ion/electron mass ratio is common in order to reduce computational costs. It is thus important to understand how the instabilities scale with $m_i/m_e$ to ensure that the reduced models do not excite unrealistic instabilities, particularly since the kink instability can potentially disrupt current sheets. In kinetic theory, it is known that the drift-kink instability growth rate is greatly reduced at higher $m_i/m_e$ \citep{daughton:1999, daughton:1998}. Fig.~\ref{fig:mime} shows how the two models scale with mass ratio. The five moment model shows an increase in growth rate with mass ratio, with the maximum growth rate being found at shorter wavelengths. In contrast, the kink mode in the ten-moment model shows a decrease in the growth rate as $m_i/m_e$ increases, with the fastest growth rate occurring at $k_y\rho_i\sim 1$, in agreement with kinetic results. The differences between these models show the importance of keeping the non-isotropic pressure tensor in modeling the kink instability. 

\subsection{Scaling with relaxation parameters}

The local ten-moment model we use has free parameters, the relaxation constants $k_{0,s}$ for the different species. In previous studies \citep{wang:2015,ng:2015,wang:2018}, it was found that setting $k_{0,s} \sim 1/d_s$ was suitable for modeling magnetic reconnection. It is thus  important to understand how the kink instabilities are affected by different $k_0$ and if the values used for reconnection are suitable for studying these instabilities.

\begin{figure}
\ig{3.375in}{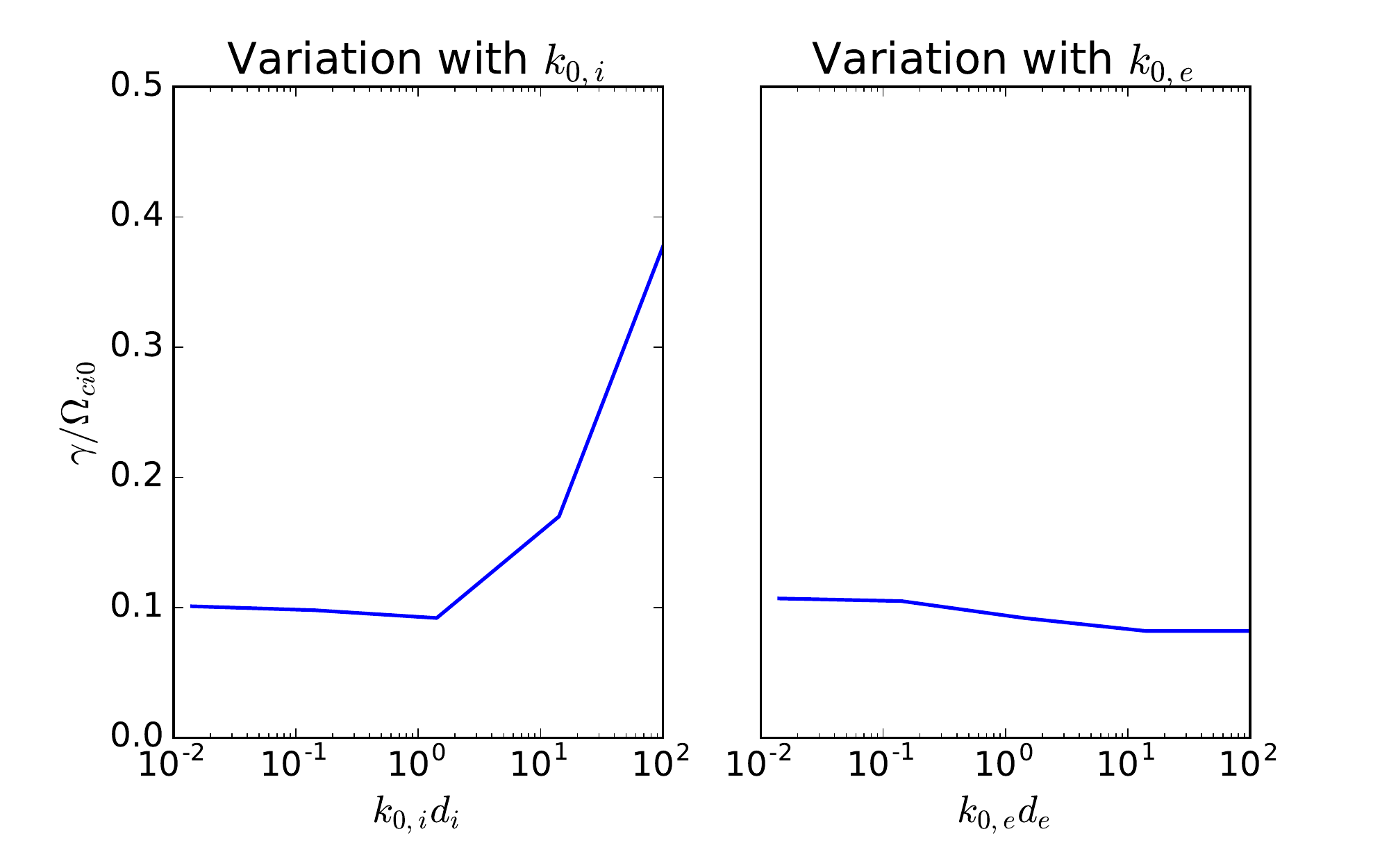}
\caption{Fastest growth rate as a function of the relaxation parameter $k_{0,s}$. }
\label{fig:kscaling}
\end{figure}

We perform two scaling studies, one in which we hold the ion relaxation parameter constant at $1/d_i$, and one in which we hold the electron parameter constant at $1/d_e$. \add{The mass ratio remains $m_i/m_e = 64$.} The results are shown in Fig.~\ref{fig:kscaling}. For the kink instability, the variation of $k_{0,i}$ has a greater effect on the maximum growth rate. As $k_{0,i}$ is increased from $1/d_i$ to $100/d_i$, the ions are isotropised and the fluid model for ions more closely resembles the five moment model, with an increase in growth rate and a shift of the fastest growing mode to longer wavelength. Decreasing the value of $k_{0,i}$ has a small impact on the growth rate, with an increase of $< 0.01\Omega_{ci0}$ over two orders of magnitude. The effect of the electron relaxation parameter is comparatively small, with an increase in growth rate at smaller $k_{0,e}$\change{,}{.} \change{indicating}{These results indicate} that retaining the additional ion physics is sufficient for describing the kink mode. 

\section{Lower hybrid drift instability} 
\label{sec:lhdi}

The equation systems \eqref{eq:fivemom} and \eqref{eq:tenmom} also support the lower hybrid drift instability (LHDI) \citep{daughton:2003,yoon:2002}. These instabilities can be found at either $k_y(\rho_i\rho_e)^{1/2} \sim 1$ or $k_y\rho_e \sim 1$. The shorter wavelength modes ($k_y\rho_e \sim 1$) have frequency on order of $\omega_{lh} \approx \sqrt{\Omega_{ci}\Omega_{ce}}$ \citep{davidson:1977,daughton:2003} and are localised around the edge of the current sheet, while the longer wavelength modes have a lower frequency and can penetrate to the centre of the current sheet \citep{daughton:2003}. 

We first review the local kinetic theory of the LHDI in order to guide our understanding of how to approximate the LHDI using fluid models. In the cold electron limit, for $k_y\rho_e \gg 1$ modes, the local dispersion relation of the LHDI can be written as \citep{davidson:1977}

\begin{equation}
1 + \frac{\omega_{pe}^2}{\omega_{ce}^2}\left(1+\frac{\omega_{pe}^2}{c^2k_y^2}\right) - \frac{2\omega_{pi}^2}{k_y^2v_{ti}^2}\left(1+\beta_i/2\right)\frac{k_y V_{di}}{\omega + k_y V_{di}} + \frac{ 2 \omega_{pi}^2}{k_y^2v_{ti}^2}\left[1+\zeta_i Z(\zeta_i)\right] = 0
\label{eq:lhdikin}
\end{equation}
where $Z(\zeta_i)$ is the ion plasma dispersion function, $\zeta_i = \omega/k_yv_{ti}$ and $V_{di}$ is the ion diamagnetic drift velocity. Note that this is the dispersion relation in the ion rest frame, so any comparisons with our Eqs~\eqref{eq:fivemom} and \eqref{eq:tenmom} should be Doppler shifted. In this limit the ion kinetic response is assumed to be unmagnetised, and gradients of perturbed quantities in the $z$ direction are neglected. 

The dispersion relation has known unstable solutions in the \change{fluid}{adiabatic} ($\zeta_i \gg 1$) and kinetic ($\zeta_i \ll 1$) \note{added subscript} limits \citep{hirose:1972}, through coupling of the drift and lower hybrid wave or the ion resonance. Because the ions can be treated as unmagnetised, the nonlocal closure of \citep{hammett:1990} or 3-d generalisation of \citep{ng:2017} would be the best fluid model for capturing the kinetic ion physics. A discussion of how well the fluid models approximate $1+\zeta_i Z(\zeta_i)$ is in the appendix.

For the electrons, since $k_y\rho_e \sim 1$, we do not expect the nonlocal closure to be applicable in this regime for the cross-field heat flux. In a more general situation, a gyrofluid model with finite Larmor radius effects would be appropriate \citep{snyder:2001,passot:2018}. However, in a reconnecting current sheet with no guide field, such models would be inapplicable close to the centre of the sheet as the magnetic field becomes close to zero. In this case, particularly in the cold electron limit where the electron dynamics affect the rate but not the instability threshold \citep{davidson:1977}, a ten-moment model with reduced isotropisation may be a better description. The role of the electron closure is discussed in Section \ref{sec:elcclosure}. 


\subsection{Results}

In this section we study the electrostatic and electromagnetic LHDI using the five moment model, the ten moment local model with the closure used in reconnection models ($k_{0,s} = 1/d_s$), the local model with $k_{0,i} = k_y$ and no electron relaxation, and the ten moment model with a nonlocal closure for ions and no electron relaxation. We use two current sheets, a thicker sheet with $\rho_i = L$ and hotter ions $T_i/T_e = 5$, where Eq.~\eqref{eq:lhdikin} would be most applicable, and a current sheet with the parameters in \citep{daughton:2003} for direct comparison with kinetic work. 

\subsubsection{Electrostatic LHDI}

We first present the results of calculations for the electrostatic LHDI. The parameters used here are $m_i/m_e = 256$, $T_i/T_e = 5$, $\omega_{pe}/\Omega_{ce} = 5$ and $\rho_i/L = 1$. Modes are calculated for $k_y \rho_e = 0.5$. \add{The current sheets support a spectrum of unstable modes} (``harmonics'' in \citep{daughton:2003}) \add{and we show four for each model and configuration we study in the figures.}

\begin{figure}
\ig{4.375in}{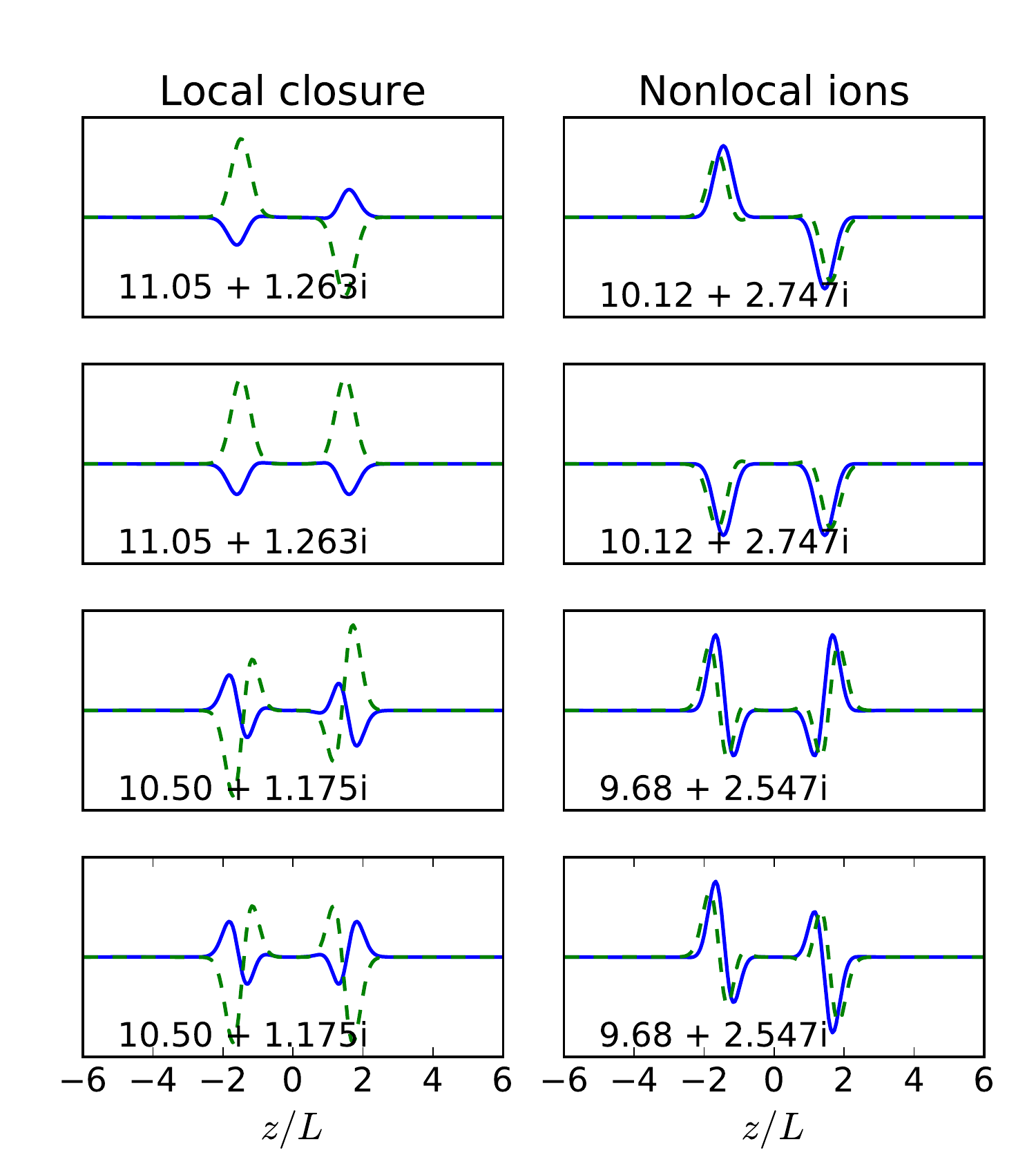}
\caption{Real (solid) and imaginary (dashed) parts of $E_y$ for the short wavelength ($k_y\rho_e = 0.5$) lower hybrid drift instability. The frequencies and growth rate are normalised to $\Omega_{ci0}$. \remove{The first column from the left uses the five-moment model.} The \change{centre}{left} column uses the ten moment model with $k_{0,e} = 0$, $k_{0,i} = k_y$ and the right column uses the nonlocal closure for ions and the local closure with $k_{0,e}=0$ for electrons. We were unable to find unstable modes for the parameters used in reconnection studies. }
\label{fig:thick_swlhdi}
\end{figure}

The eigenmodes are shown in Fig.~\ref{fig:thick_swlhdi}, where we use the \remove{five moment model,} the local ten moment model with $k_{0,e} = 0, k_{0,i}=k_y$ and the ten moment model with nonlocal ions and $k_{0,e} = 0$. For this set of parameters, the \add{five-moment model and} ten\add{-}moment model with $k_{0,s} = 1/d_s$ \change{was}{are} stable to lower hybrid instabilities. 

\change{In this instance, the five moment model is comparatively stable, with the growth rate being $5$ to $10$ times smaller than calculated using the nonlocal ten moment model. This}{In this case, the stability of the five-moment model} is likely due to its inability to model the ion response correctly, which \change{are be}{is} important in thicker sheets with a smaller equilibrium drift velocity \citep{davidson:1977}. Both ten moment calculations show that the LHDI is present, and the model using the nonlocal closure for the ions has a growth rate and structure that is consistent with local kinetic theory, which gives a growth rate of $\gamma/\Omega_{ci0} = 2.9$ in the region around $z/L = 1.5$.

\begin{figure}
\ig{3.375in}{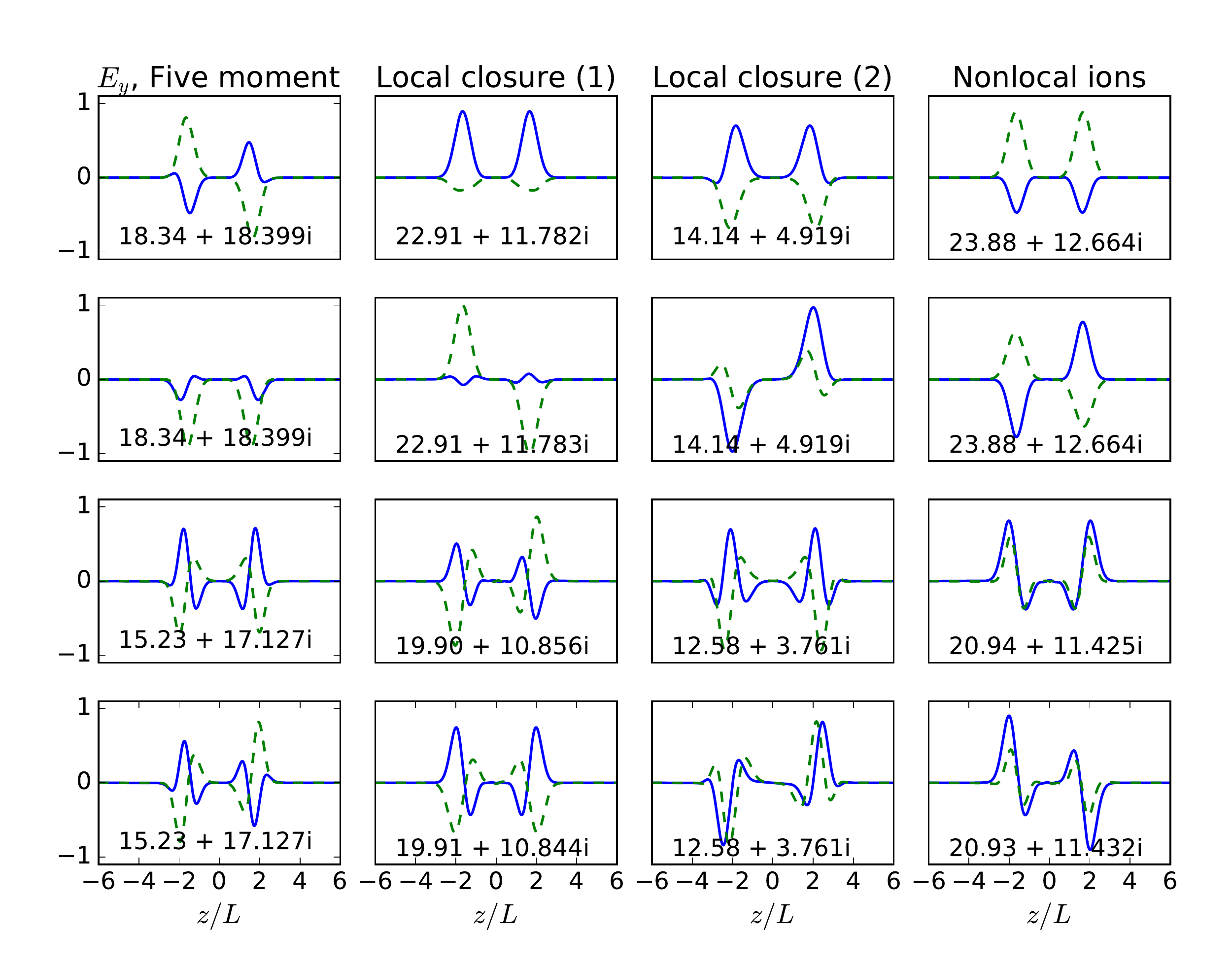}
\caption{Real (solid) and imaginary (dashed) parts of $E_y$ for the short wavelength ($k_y\rho_e = 1$) lower hybrid drift instability. The frequencies and growth rate are normalised to $\Omega_{ci0}$. The first column from the left uses the five-moment model. The second column (labeled (1)) uses the ten moment model with $k_{0,e} = 0$, $k_{0,i} = k_y$, the third column (labeled (2)) uses $k_{0,s} = 1/d_s$ and the final column uses the nonlocal closure for ions and the local closure with $k_{0,e}=0$ for electrons. }
\label{fig:thin_swlhdi}
\end{figure}

Figure~\ref{fig:thin_swlhdi} shows the eigenmode calculations for the second set of parameters, with $\rho_i/L = 2$, $T_i = T_e$, $m_i/m_e = 512$ and $\omega_{pe}/\omega_{ce} = 5$, also used in \citep{daughton:2003}. This is a comparatively thinner current sheet, and all the models are unstable to the LHDI, with similar mode structures but different growth rates. The five moment model has the fastest growing modes, while the ten moment models with $k_{0,e} = 0$ have growth rates and frequencies differing by less than $\Omega_{ci0}$. For similar real frequency, these modes have a faster growth rate than in the kinetic calculation of \citep{daughton:2003}. The model with $k_{0,s} = 1/d_s$ shows the lowest growth rates, which is a general trend reproduced in the next sections. 

\subsubsection{Electromagnetic LHDI}

\begin{figure}
\ig{4.375in}{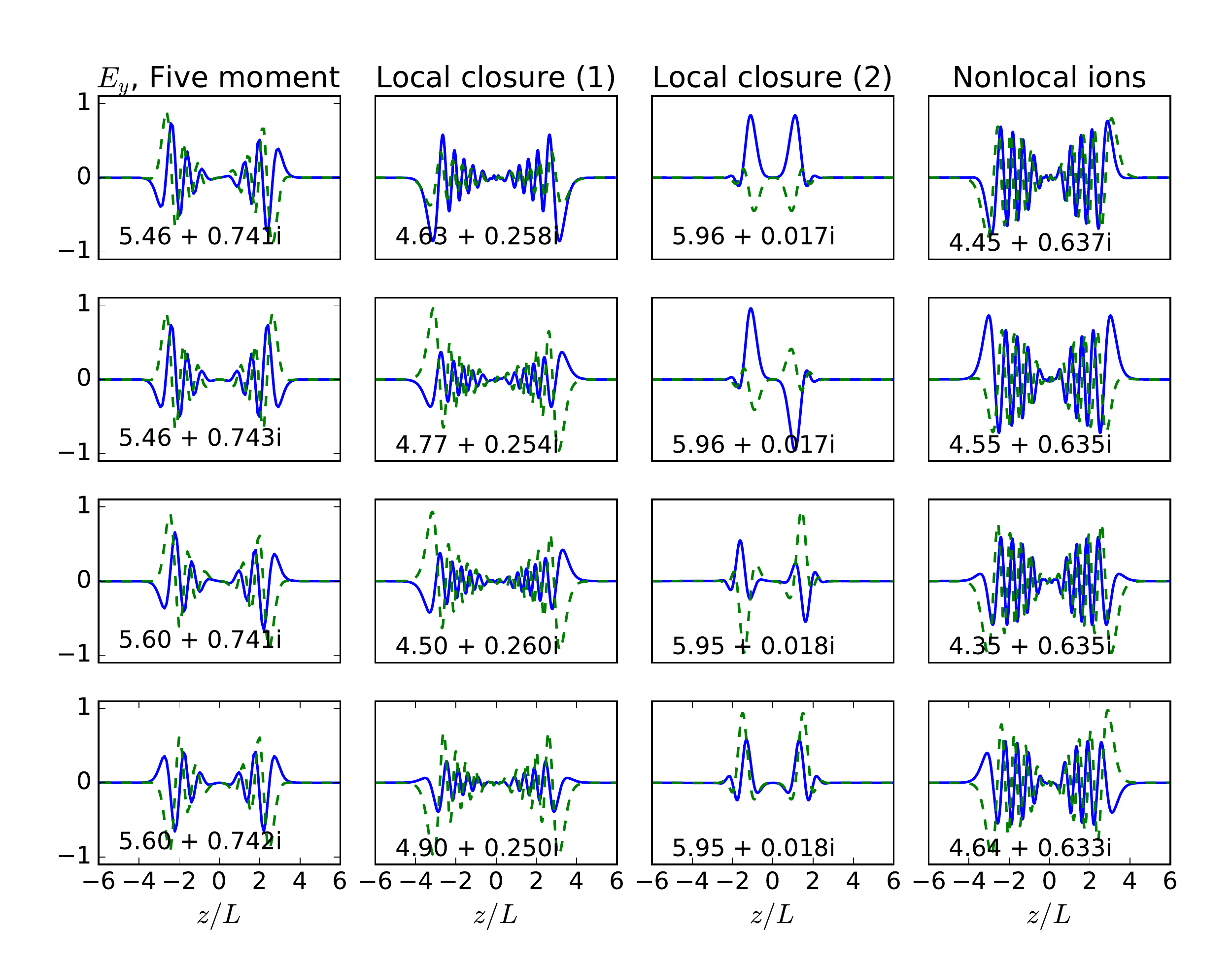}
\caption{Real (solid) and imaginary (dashed) parts of $E_y$ for the long wavelength ($k_y\sqrt{\rho_i\rho_e} = 1$) lower hybrid drift instability. The frequencies and growth rate are normalised to $\Omega_{ci0}$. The first column from the left uses the five-moment model. The second column (labeled (1)) uses the ten moment model with $k_{0,e} = 0$, $k_{0,i} = k_y$, the third column (labeled (2)) uses $k_{0,s} = 1/d_s$ and the final column uses the nonlocal closure for ions and the local closure with $k_{0,e}=0$ for electrons.}
\label{fig:thick_lwlhdi}
\end{figure}

For the longer wavelength electromagnetic LHDI, we first present the results of calculations with the thicker $\rho_i = L$ current sheet with the same parameters as the previous \change{second}{section}, but wavelength $k_y\sqrt{\rho_i\rho_e} = 1$. Selected eigenmodes are shown in Fig.~\ref{fig:thick_lwlhdi}, where the local model labeled (1) again refers to the closure with $k_{0,e}= 0$, $k_{0,i} = k_y$, and the model labeled (2) has $k_{0,s} = 1/d_s$. Here the local model with $k_{0,s} = 1/d_s$ shows very weakly unstable modes, with structures reminiscent of the electrostatic modes, while the five moment model and other ten moment models show broader mode structures which extend to the centre of the current sheet, which is expected for these modes \citep{daughton:2003}. 

\begin{figure}
\ig{4.375in}{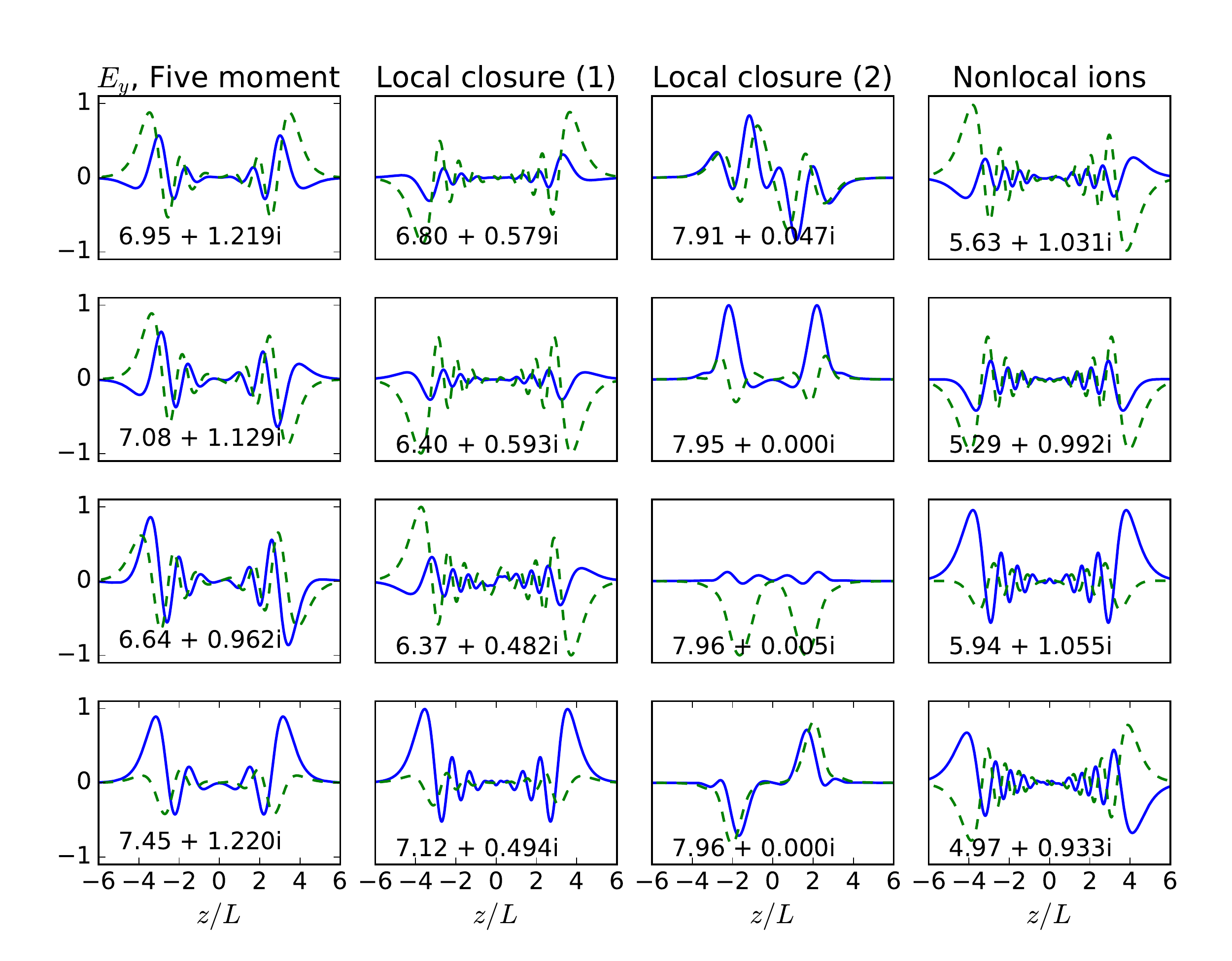}
\caption{Real (solid) and imaginary (dashed) parts of $E_y$ for the long wavelength ($k_y\sqrt{\rho_i\rho_e} \approx 0.84$) lower hybrid drift instability. The frequencies and growth rate are normalised to $\Omega_{ci0}$. The first column from the left uses the five-moment model. The second column (labeled (1)) uses the ten moment model with $k_{0,e} = 0$, $k_{0,i} = k_y$, the third column (labeled (2)) uses $k_{0,s} = 1/d_s$ and the final column uses the nonlocal closure for ions and the local closure with $k_{0,e}=0$ for electrons.}
\label{fig:thin_lwlhdi}
\end{figure}

With the thinner sheet used in \citep{daughton:2003}, we find the modes shown in Fig.~\ref{fig:thin_lwlhdi}. Again, the ten moment model with the parameters used in reconnection studies ($k_{0,s} = 1/d_s$) shows much more stable modes around the drift frequency $k_y u_i \approx 8.0 \Omega_{ci0}$. The five moment model, ten moment model with nonlocal ions and with $k_{0,e} =0, k_{0,i}=k_y$ are able to capture the electromagnetic LHDI, though the growth rates show quantitative differences with the results of \citep{daughton:2003}, though the frequencies and growth rates have a similar range of values. We believe the discrepancies are due to the limitations of our electron model, which will be demonstrated in the next section. 

\subsubsection{Sensitivity to electron model}
\label{sec:elcclosure}

Although the ten-moment model contains non-gyrotropic pressure effects, it is not clear how this affects the calculations of the lower hybrid instabilities where $k_y \rho_e \sim 1$. As we did with the kink instability, we perform a scaling of the electron relaxation parameter and study how the fastest growing lower hybrid mode varies with $k_{0,e}$. In these calculations we use the nonlocal ten moment ion model as it is the best approximation to the ion kinetic response. 

\begin{figure}
\ig{4.375in}{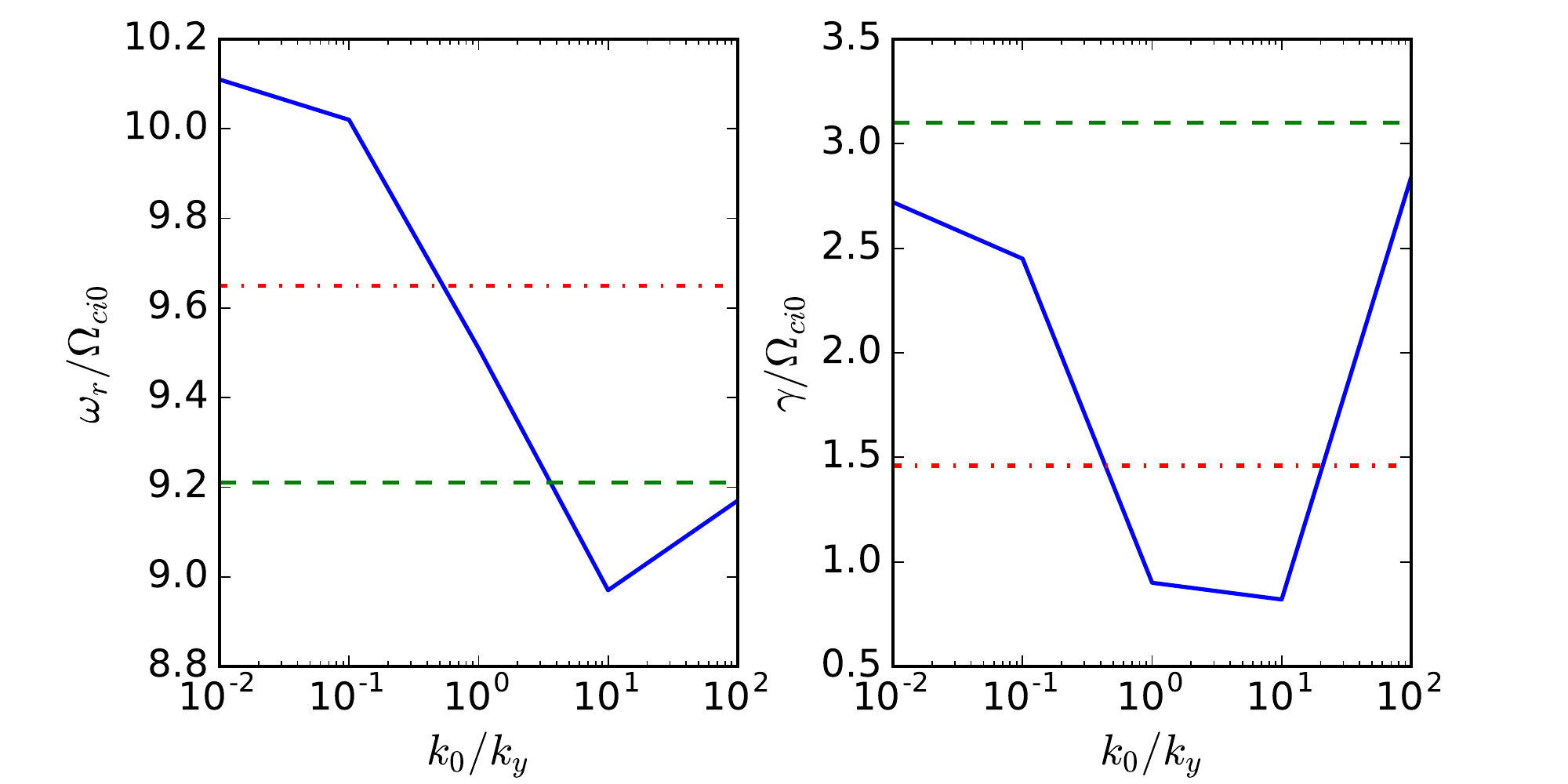}
\caption{Frequency and growth rate of the fastest growing lower hybrid mode as $k_{0,e}$ is varied while using the nonlocal model for ions. The dashed line shows the growth rate when using the five moment model for electrons, while the dash-dotted line uses the nonlocal model for both ions and electrons. }
\label{fig:lhdi-scaling}
\end{figure}

The parameters of the current sheet used in this scan are $m_i/m_e = 256$, $T_i/T_e = 5$, $\omega_{pe}/\Omega_{ce} = 5$ and $\rho_i/L = 1$, and we use $k_y\rho_e = 0.5$. The results are shown in Fig.~\ref{fig:lhdi-scaling}, with calculations using nonlocal and five moment electrons also plotted for reference. The growth rate of the instability is larger in the limits $k_0 \to 0$ and $k_0 \to \infty$, and has a minimum for intermediate values of $k_0$.  

In the limit of $k_0 \gg k_y$, the electrons are isotropised, so that the ten moment model approaches the five moment limit, while as $k_0 \to 0$, the pressure tensor is allowed to evolve freely. The calculation with nonlocal electrons has growth rate close to the case with $k_0=k_y$, which is consistent with the $k_y$ dependence of the nonlocal heat flux. 

\add{ Which closure approximation best captures the LHDI remains a question for further study. As mentioned earlier, Equation } \eqref{eq:lhdikin} \add{is the local dispersion relation in the cold electron limit, where finite Larmor radius effects can be neglected. For arbitrary $T_e$, however, it is still necessary to incorporate these effects} \citep{davidson:1977} \add{. Outside the current sheet, electrons are strongly magnetised, so a gyrofluid closure with finite Larmor radius approximations would work ($k_y = k_\perp$ for this geometry) } \citep{snyder:2001,passot:2018}, \add{while within the current sheet, where electrons are unmagnetised, the nonlocal closure discussed above would correctly capture the electron response. A transition between the two limits will thus be necessary to capture the instability correctly in the ten-moment model.}

\remove{In order to properly describe the electron dynamics, a more complex closure with spatial dependence will be required as the electrons are magnetised outside the current sheet while being unmagnetised within the sheet. }

\subsection{Simulations of the LHDI}
\label{sec:simulations}
\note{changed to section 5.2}

The five and ten moment equation systems have been implemented in the finite-volume version of the \texttt{Gkeyll} code, which uses a high-resolution wave propagation method for the hyperbolic part of the equations and a point implicit method for the source terms \citep{hakim:2006,hakim:2008}, and has previously been used to study magnetic reconnection \citep{wang:2015, ng:2015, ng:2017}. The kinetic simulations use the discontinuous Galerkin \add{finite-element} Vlasov-Maxwell solver of \texttt{Gkeyll 2.0} \citep{juno:2018}. \add{Because the Vlasov code uses a discontinuous Galerkin method, we require a basis function expansion in each cell, and we choose piecewise quadratic basis functions from the Serendipity Element family. Details on the particulars of the basis expansion can be found in} \citep{arnold:2011,juno:2018}.

The simulations presented below use the parameters $\rho_i = L$, $v_{t,e} = 0.06$, $m_i/m_e = 36, T_i/T_e = 10$, with simulation domain $L_y\times L_z = 6.4L\times 12.8 L$. A background plasma with $n_b = 0.001 n_0$ is introduced for numerical stability. In the fluid simulations the grid size was $N_x\times N_y = 256\times 512$. The kinetic simulations are run in two velocity dimensions (2X2V) as the LHDI (with no guide field and $k=k_y$) does not depend on the out-of-plane velocity. The configuration space dimensions are the same as the fluid simulations, but use $96\times 192$ cells, with quadratic Serendipity elements \citep{arnold:2011}\note{changed cite to citep} in each cell. The electron velocity domain ranges from $-8 v_{t,e}$ to $8 v_{t,e}$ while the ion velocity domain ranges from $-6 v_{t,i}$ to $6 v_{t,i}$ in each direction. Two cells with quadratic serendipity elements are used per species thermal velocity. Quadratic Serendipity elements give us roughly a factor of 3 in additional sub-cell resolution for a total amount of resolution of 2.3 grid points per $\rho_e$ (calculated using the asymptotic $B$ field) and 6 cells per thermal velocity. 

These parameters were chosen to balance computational costs while maintaining $T_i \gg T_e$ and the local approximation where gradients of the perturbed quantities in the $y$ direction are much larger than gradients in the $z$ direction. In the simulations, an initial $m=8$ perturbation is imposed, which corresponds to $k_yL \approx 7.9$ or $k_y\rho_e \approx 0.41$ and is close to the wavelength with the maximum growth rate predicted by local theory \citep{davidson:1977}. For these parameters, the predicted kinetic growth rate is $\gamma=1.1\Omega_{ci0}$ and the most unstable region is at $z/L \approx 1.6$. \add{In order to compare the (2X2V) simulations to the fluid simulations, we use an adiabatic index of 2 for the five-moment simulations, and modify the ten-moment model to relax only the in-plane components of the pressure tensor when using the local closure. We do not modify the nonlocal model as the closure does not couple the out-of-plane diagonal component of the pressure tensor to the in-plane components.} 

\begin{figure}
\ig{4.375in}{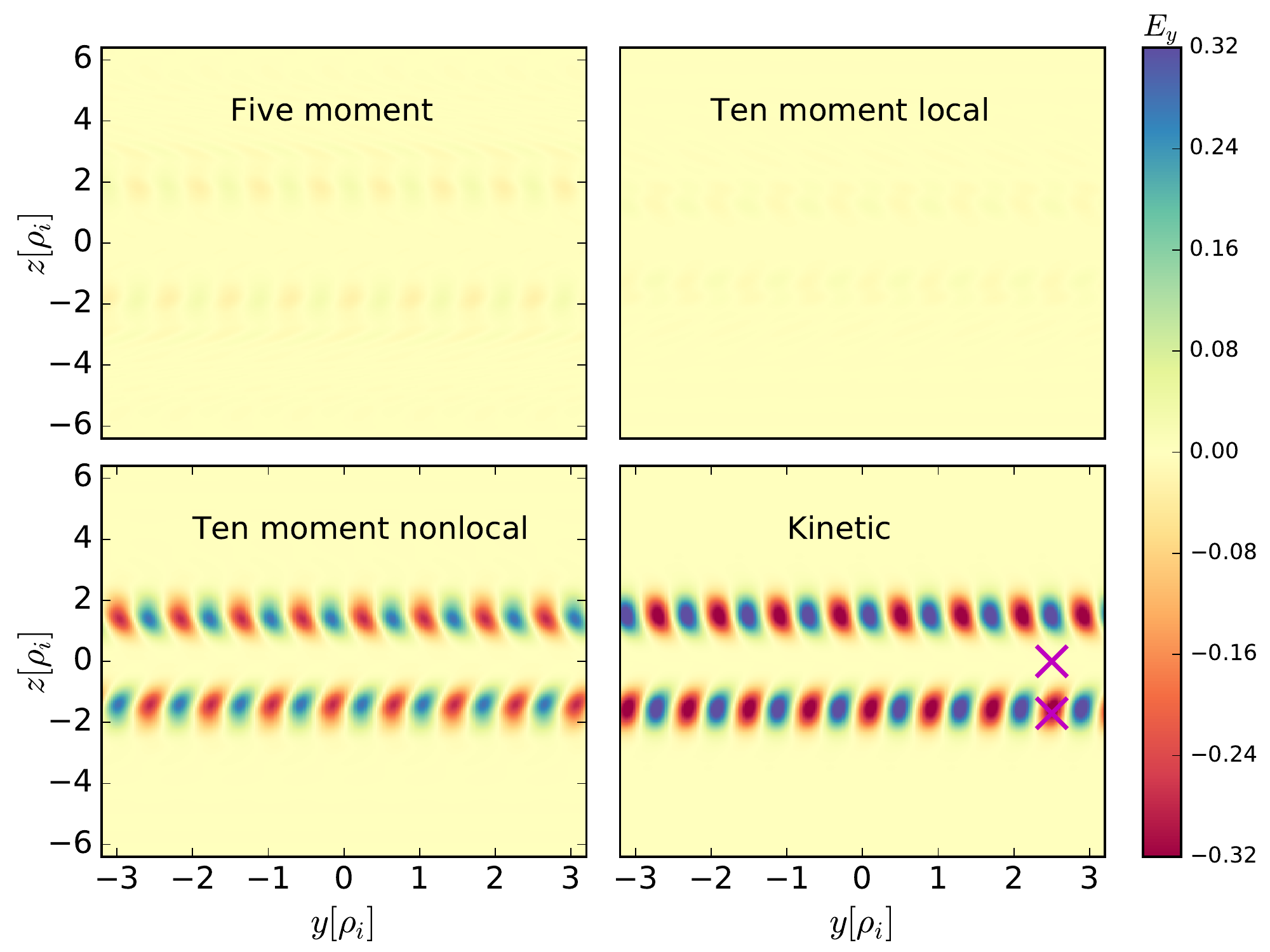}
\caption{Structure of $E_y$ in simulations of the LHDI using different models at $t\Omega_{ci} = 6$. $E_y$ is normalised to $B_0v_{A0}$.}
\label{fig:fields}
\end{figure}

Fig.~\ref{fig:fields} shows a comparison between the structure of $E_y$ in kinetic and fluid simulations of the LHDI at $t\Omega_{ci0} = 6$. The local closure used $k_{0,i} = k_y$ and $k_{0,e}= 0$. The calculation with the nonlocal closure for ions also used $k_{0,e}=0$. From the simulations, the measured growth rates of the $m=8$ mode were \change{$0.39\Omega_{ci0}$}{$0.34\Omega_{ci0}$} for the local closure, $0.84\Omega_{ci0}$ for the nonlocal closure and $1.1\Omega_{ci0}$ for the kinetic model. The mode was found to be stable when using the five moment model. As can be seen in the lower panels, the structure of the LHDI in this regime is well described by the nonlocal model, with the growth rate about $24\%$ slower than in the kinetic simulation. When using the local model, in spite of the slower growth rate, the LHDI does eventually develop with a similar structure. \add{When performing the equivalent fluid simulations with three velocity dimensions -- with an adiabatic index of $5/3$ or relaxing all the components of the pressure tensor -- which would be used in simulations of physical systems, the results are similar. The five-moment model is stable for these parameters, while the local relaxation shows a small increase in the growth rate to $0.39\Omega_{ci0}$}.

\begin{figure}
\ig{4.375in}{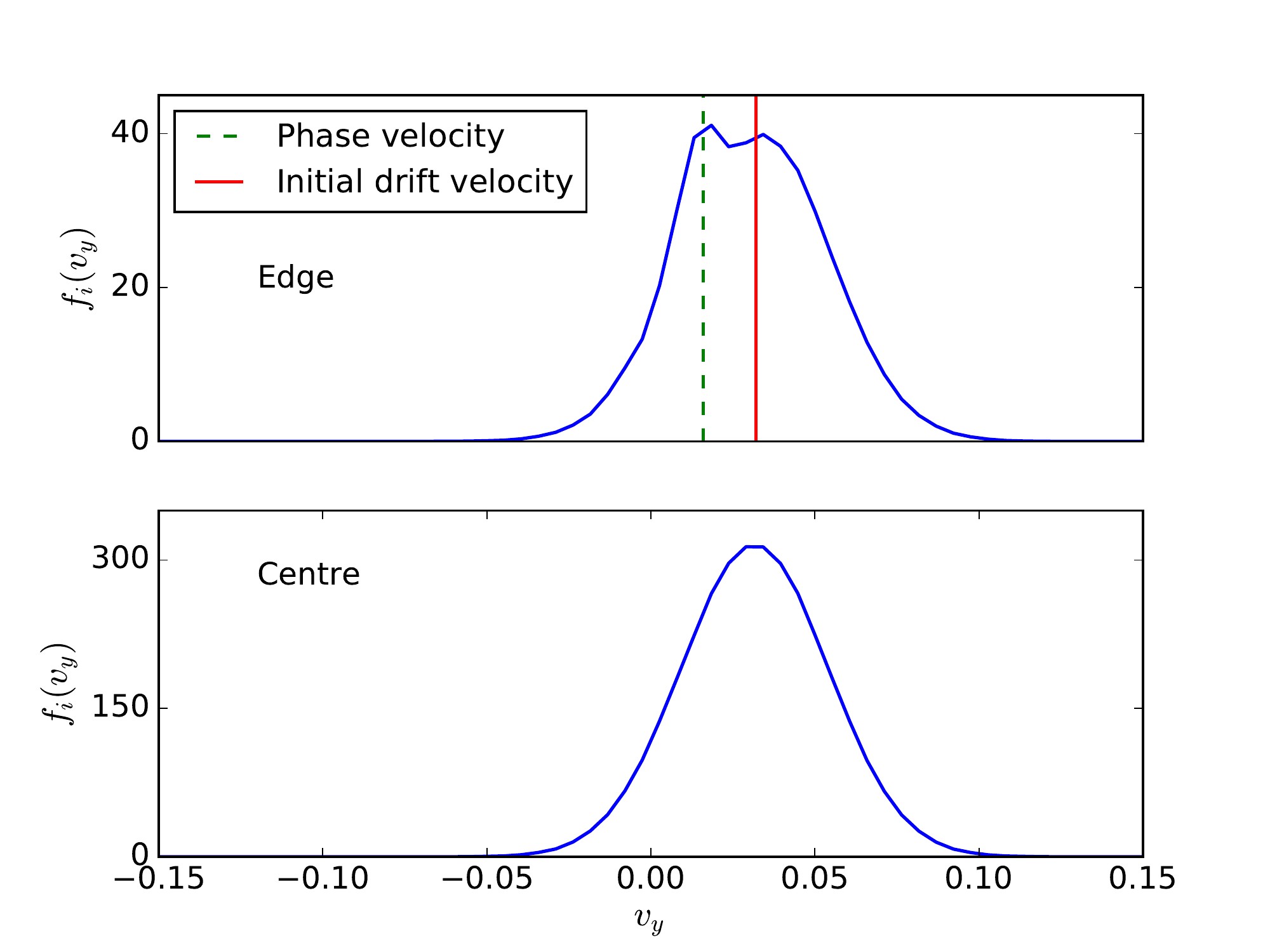}
\caption{Ion distribution functions $f(v_y,v_z=0)$ at $z = 0$, at the centre of the current sheet and $z = -1.7$, at the edge of the sheet (magenta crosses in Fig.~\ref{fig:fields}). }
\label{fig:distfunc}
\end{figure}

The role of ion kinetic effects is highlighted in Fig.~\ref{fig:distfunc}, which shows a cut of the ion distribution function $f(v_y,v_z=0)$ at the edge, where the perturbed electric field is confined to, and centre of the current sheet. These points are marked in Fig.~\ref{fig:fields}. At the edge of the current sheet, the ion resonance can be seen in the upper panel, where the phase velocity from the theoretical solution and the initial drift velocity are marked. At the centre of the sheet, the distribution remains close to Maxwellian, consistent with the electrostatic LHDI being confined to the edge of the current sheet. 

\section{Conclusion}
\label{sec:conclusion}

We have performed calculations of the drift-kink and lower hybrid drift instabilities for Harris sheets using the five and ten moment two-fluid models. For the drift-kink instability, the ten-moment model has growth rates and wavenumbers comparable to the results of Vlasov-Maxwell calculations, unlike the five-moment, or standard two-fluid, model, which has faster growing modes at larger mass ratios and wavenumbers. The growth rates are not sensitive to the relaxation parameter in the range $k_{0,s}\lesssim 1/d_s$. Additionally, the sausage moment is damped by the ten-moment model, which is consistent with kinetic studies \cite{pritchett:1996kink,daughton:1999}. Although the kink mode has a lower growth rate at high mass ratio, this result does not preclude its excitation as a secondary instability \citep{lapenta:2002}, or the growth of ion-ion kink instabilities in the ten moment model \citep{karimabadi:2003a,karimabadi:2003b}, which could be the topic of future work.

The results are consistent the fluid work of \citep{pritchett:1996,daughton:1999kink} in the long wavelength regime, and the scaling with physical parameters such as temperature ratio and mass ratio are consistent with kinetic models \citep{daughton:1999}. There are some differences compared to the fluid results of \citep{yoon:2002} due to the treatment of the pressure term in the momentum equation. The kink modes are more sensitive to the ion model used, which may be useful if reduced electron models are used to save on computational costs.

\add{In global simulations, the importance of using a closure that captures the kink mode correctly has been seen in global simulations of Ganymede} \citep{wang:2018, ng:thesis}. In \citet{ng:thesis}, \add{it was shown that when using a five-moment model to study magnetosphere dynamics, a kink instability was excited in the magnetotail after the formation of the tail current sheet. While the current sheet did not disrupt in this case, the instability caused the formation of a large-scale corrugated structure. In the ten-moment simulations} \citep{wang:2018, ng:thesis}, \add{which are expected to show better agreement with kinetic models for this instability, the growth of the kink instability was much reduced. } 

The LHDI can be observed in both five and ten moment models with the appropriate choice of closure. Based on the results of kinetic theory \citep{davidson:1977,hirose:1972}, it is clear that the ions should be modeled using a nonlocal closure, or a relaxation with the $k_{0,i} \approx k_y$ so that the ion resonance can be captured. However, there is sensitivity to the electron model used, and the instabilities have the largest growth rates when $k_{0,e} \to 0$ or $\infty$. For parameters used in reconnection studies ($k_{0,s} = 1/d_s$) \citep{wang:2015,ng:2015}, both electromangetic and electrostatic instabilities are damped, with the electromagnetic modes being stable in some regimes. 

Finally, we have performed comparisons of fluid and Vlasov-Maxwell simulations to show that the LHDI can be observed in our fluid simulations. The distribution function information demonstrates the importance of the ion resonance and applicability of the local kinetic theory for thicker sheets, and illustrates the utility of the Vlasov-Maxwell code in analysing distributions due to the lack of particle noise as compared to particle-in-cell simulations.

In the context of global simulations, where we would want to model magnetic reconnection in addition to these instabilities, it may not be possible to capture the kink, LHDI and reconnection simultaneously in certain regimes. Due to computational constraints, it would be very computationally intensive to use the nonlocal closure for the ions in global studies, and setting $k_{0,i}=k_y$  to capture the LHDI would only be appropriate for a small range of wavenumbers, and likely excite the drift-kink unphysically. Additionally, the LHDI has a reduced growth rate when using closure parameters similar to those in reconnection studies, though this may not be an issue for the electrostatic LHDI in sufficiently thin sheets. A compromise may be the use of the five moment model for electrons, though this would miss the electron pressure tensor effects on reconnection. Even with compromises, it is evident that the ten moment model is a significant improvement over MHD models and captures some key physics features of fully kinetic simulations, which cannot be used for global space weather studies.

There is potential for further development of the closure with the use of temperature gradients, which provides some heat flux while remaining computationally tractable \citep{allmann:2018}\note{changed cite to citep}, but more study on how this model affects reconnection and instabilities is required. With the current ten-moment model, the drift-kink instability and reconnection can be studied simultaneously, which avoids the unphysical growth of the kink mode in five moment models that can disrupt the current sheet.

\acknowledgments

J.~Ng, A.~Hakim and A.~Bhattacharjee are supported by NSF Grant AGS-1338944 and DOE contract DE-AC02-09CH11466. The work of J. Juno was supported by a NASA Earth and Space Science Fellowship, Grant no. 80NSSC17K0428. This research used resources of the National Energy Research Scientific Computing Center, a DOE Office of Science User Facility supported by the Office of Science of the U.S. Department of Energy under Contract No. DE-AC02-05CH11231. Data for the figures are available online at \citep{ng:2018data}. \add{We are grateful to J.~TenBarge for valuable discussions.} The calculations in this work made use of the \texttt{scipy} library \citep{scipy}.

\appendix

\section{Electrostatic response in the various plasma models}

How well the fluid models approximate the ion response is determined by the term proportional to $R(\zeta_i) = 1+\zeta_i Z(\zeta_i)$ in \eqref{eq:lhdikin}. This can be calculated by solving the 1-D dispersion relation in the fluid models and finding the perturbed density \citep{hammett:1990}
\begin{equation}
n_1 = -n_0 \frac{e\phi_1}{T_0} R(\zeta)
\end{equation}
where $\phi_1$ is the perturbed electrostatic potential. 

In the five moment model with no thermal conductivity or viscosity, the response can be written as 
\begin{equation}
R_{5mom}(\zeta) = \frac{1}{\gamma - 2\zeta^2}
\end{equation}
where $\gamma$ is the adiabatic constant (we use $5/3$ in this paper).

The nonlocal ten-moment model has \citep{hammett:1990}
\begin{equation}
R_{nonlocal}(\zeta) = \frac{\chi_1-i\zeta}{\chi_1 - 3i\zeta - 2\chi_1\zeta^2 + 2 i \zeta^3},
\end{equation}
with $\chi_1 = 2/\sqrt{\pi}$, and the relaxation to local isotropy has
\begin{equation}
R_{local}(\zeta) = \frac{\alpha - i \zeta}{5\alpha/3 - 3 i \zeta - 2 \alpha \zeta^2 + 2 i \zeta^3}
\end{equation}
where $\alpha = k_0/k$, with $k_0 v_t$ being the relaxation rate. The dependence of the response function on the relaxation parameter is quite explicit and shows why the choice of $k_0$ is important the regime where the ion resonance is important for the LHDI. There is also a subtle difference between relaxing the pressure to local isotropy and relaxing temperature fluctuations to the equilibrium temperature $T_0$, which is responsible for the $5/3$ in the first term of the denominator. 


\end{document}